\documentclass[twocolumn,showpacs,pra]{revtex4}
\usepackage{amsfonts}
\usepackage{amsmath}
\usepackage{amssymb}
\usepackage{graphicx}

\begin{document}

\title{Hamiltonian chaos in a coupled BEC -- optomechanical cavity system}
\author{K. Zhang$^{1}$}
\author{W. Chen$^{2}$}
\author{M. Bhattacharya$^{2}$}
\author{P. Meystre$^{2}$}
\affiliation{\begin{tabular}{c}
$^{1}$State Key Laboratory of Precision Spectroscopy, Dept. of Physics, East
China Normal University, Shanghai 200062, China \\
$^{2}$B2 Institute, Dept. of Physics and College of Optical Sciences, The
University of Arizona, Tucson, AZ 85721, USA%
\end{tabular}}


\begin{abstract}
We study a hybrid optomechanical system consisting of a Bose-Einstein
condensate (BEC) trapped inside a single-mode optical cavity with a moving
end-mirror. The intracavity light field has a dual role: it excites a
momentum side-mode of the condensate, and acts as a nonlinear spring that
couples the vibrating mirror to that collective density excitation. We
present the dynamics in a regime where the intracavity optical field, the
mirror, and the side-mode excitation all display bistable behavior. In this
regime we find that the dynamics of the system exhibits Hamiltonian chaos
for appropriate initial conditions.
\end{abstract}

\maketitle

\section{Introduction}
Cavity optomechanics is currently the focus of extensive theoretical and
experimental investigations and has witnessed spectacular advances in the
last few years, see Ref.~\cite{Vahala2008} and references therein. A central
paradigm of the field has been the cooling of one or several modes of
vibration of a movable mirror in an optical cavity to their quantum
mechanical ground state. Rapid progress has been made towards this goal
using micro- and nano-fabricated mirrors \cite{coolm}, membranes \cite%
{Thompson2008} and zippers \cite{Painter2009}. Equally impressive has been
the demonstration of optomechanics effects in situations where the
oscillating mirror is replaced by other physical objects, most notably
perhaps an ultracold atomic gas \cite{Kurn2008} or a BEC \cite{Esslinger}.
These experiments are first steps in opening another frontier broached by
optomechanics - the coherent coupling of mechanical elements with atomic
systems \cite{Hansch07,Sun08}, or more generally the interfacing of solid state
physics with AMO science. It is expected that these developments will lead
to basic advances, for instance in the understanding of the
quantum-classical interface, as well as to novel applications related to
measurement and sensing \cite{Giscard2009}, coherent control
\cite{Hansch07,Singh2008} and quantum information processing \cite{Rabl2009}.

\begin{figure}[tbp]
\includegraphics[width=6cm]{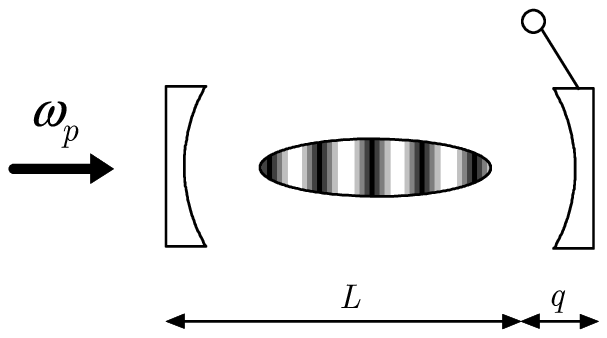}
\caption{Bose-Einstein condensate trapped inside a Fabry-P{\' e}rot cavity
with a moving end-mirror. Due to photon recoil the intracavity light field
induces a matter-wave momentum side-mode in the condensate that is formally
analogous to a second, microscopic moving mirror coupled to the ``true''
moving mirror via the optical field.}
\label{cavity}
\end{figure}

The present paper considers theoretically a system at the boundary between
cavity QED, nanoscience, and ultracold science -- a BEC trapped inside a
single-mode optical cavity with a moving mirror, see Fig.~\ref{cavity}.
This system was previously discussed in the limit where the coupling between
the BEC and the optical field is weak, the object of that study being the
ensuing bistability of the BEC Mott insulator-superfluid transition \cite%
{Chen2009}. In contrast, we now consider a situation where the BEC is
strongly coupled to the intracavity field.

It has been shown experimentally for the case of a cavity with fixed
end-mirrors that the recoil resulting from the interaction between the
ultracold atoms and the intracavity field results in the excitation of a
condensate side-mode that is (in the simplest case) formally identical with
a mechanical oscillator \cite{Esslinger}. Extending these results to the
case of a resonator with a moving end-mirror, and therefore coupling a
``microscopic mirror'' to a nanoscale device, is particularly interesting 
in view of the high level of experimental control achievable in this system.
As such it provides a test-bed for a number of investigations ranging from the
use of a BEC as a quantum sensor to characterize and control the state
of the moving mirror, \cite{Hansch07} to (conversely) the use of the moving mirror to manipulate the condensate \cite{Bhatt}, and to fundamental studies of the entanglement between macroscopic and microscopic objects \cite{Genes08} including the effects of decoherence.

At the simplest level the moving mirror and the condensate side-mode can be approximated as two harmonic oscillators coupled by the intracavity optical field, which acts as a nonlinear spring. In general, all three components of the system can exhibit a multistable behavior, which normally reduces to a bistable behavior for small mirror displacements. In the bad cavity limit that we consider here, where the optical field decay rate is faster than the decay rates of the mirrors, this bistable behavior can result in rich mirror dynamics. In particular, in the limit where the damping of the mirrors can be completely neglected, the mirrors can undergo Hamiltonian chaos for appropriate initial conditions \cite{note}. We find an energy-dependent order to chaos transition similar to that seen in the double pendulum and in the H{\' e}non-Heiles system \cite{Hilborn}.

The remainder of this paper is organized as follows: Section II introduces our model, and discusses the analogy between the condensate side-mode and a moving mirror driven by radiation pressure. We then eliminate adiabatically the intracavity field to derive coupled equations that describe the dynamics of the coupled condensate-moving mirror system in the absence of dissipation. Section III discusses the dynamics of the system, and shows the appearance of Hamiltonian chaos for appropriate initial conditions. We show Poincar{\' e} surfaces of section and spectra that illustrate that point. Finally, section IV is a summary and conclusion.

\section{The model}

We model the combined mirror-condensate-optical field system via the
Hamiltonian
\begin{equation}
\label{H}
{\hat H} = {\hat H}_c+ {\hat H}_a + {\hat H}_d,
\end{equation}
where ${\hat H}_c$ describes the intracavity field and its coupling to the
moving mirror, ${\hat H}_a$ describes the condensate and its coupling to the
light field, and ${\hat H}_d$ accounts for the coupling of the various
subsystems to thermal reservoirs and hence to dissipation.

The Hamiltonian ${\hat{H}}_{c}$ is given explicitly by \cite{Law95}
\begin{equation}
\hat{H}_{c}=-\hbar \Delta _{c}\hat{a}^{\dagger }\hat{a}-\hbar \xi \hat{a}
^{\dagger }\hat{a}\hat{q}+\frac{\hbar \omega _{m}}{2}(\hat{p}^{2}+\hat{q}
^{2})-i\hbar \eta \left( \hat{a}-\hat{a}^{\dagger }\right) ,
\end{equation}%
where $\hat{q}$ is the dimensionless position operator for the quantized
mode of vibration of the mirror of effective mass $m$ under consideration and $\hat{p}$ its momentum operator, with commutation relation $[\hat{q},\hat{p} ]=i $, $\hat{a}$ and ${\hat a}^\dagger$ are the bosonic annihilation and creation operators for the optical cavity field of frequency $\omega _{p}$, with $[a,a^{\dagger }]=1$, $\xi =\omega_{c}\sqrt{\hbar /m\omega _{m}}/L$ is the optomechanical coupling rate, and $\Delta _{c}=\omega _{p}-\omega _{c}$ is the cavity-pump detuning. Finally, the Hamiltonian $-i\hbar \eta ({\hat{a}}-{\hat{a}}^{\dagger })$ accounts for the external pumping of the cavity mode.

We describe the motion of the BEC trapped in the optomechanical cavity with a
simple one-dimensional model in which the atomic motion along the cavity
axis is quantized. For large light-atom detunings $\Delta_a$ we can
adiabatically eliminate the internal excited state dynamics of the atoms and
ignore the effects of spontaneous emission. Considering in addition low
enough atomic densities that two-body interactions can be neglected we have
\begin{equation}  \label{eq:HBEC}
\hat{H}_{a}=\int \hat{\Psi}^{\dagger }\left( x\right) \left( -\frac{\hbar
^{2}}{2m_{a}}\frac{d^{2}}{dx^{2}}+\hbar U_{0}\hat{a}^{\dagger }\hat{a}\cos
^{2}kx\right) \hat{\Psi}\left( x\right) dx
\end{equation}
where $\hat{\Psi}\left( x\right) $ is a bosonic field annihilation operator,
$m_a$ is the atomic mass, $U_0=g_{0}^{2}/\Delta_{a}$ is the far off-resonant
vacuum Rabi frequency and $k=\omega_p/c$ is the wave number of the light
field. We consider in the following the strong coupling regime $%
Ng_{0}^{2}/\left\vert \Delta _{a}\right\vert \gg \kappa $ \cite{Ritsch00},
where $N$ is the average number of atoms and $\kappa$ is the decay rate of the cavity, in which case the cavity field is
strongly dependent on the collective density distribution of the BEC. More
specifically the back-action of the BEC and the mirror can change the photon
number inside the cavity significantly (but without changing its spatial structure). (Note that we ignore for now the effect of the harmonic trap confining the atoms inside the Fabry-P{\' e}rot in Hamiltonian (\ref{eq:HBEC}).)

The photon recoil associated with the absorption and stimulated emission of
light by the BEC results in the generation of a symmetric momentum side-mode
at $\pm 2\hbar k$ \cite{Esslinger}. To account for this effect we expand the
field operator in Eq.~(\ref{eq:HBEC}) as
\begin{equation}
\hat{\Psi}\left( x\right) \simeq \left( \hat{c}_{0}+\sqrt{2}\cos (2kx)\hat{c}
_{2}\right) /\sqrt{L},
\end{equation}
where $\hat{c}_{0}$ and $\hat{c}_{2}$ are bosonic annihilation operators for
atoms in the zero-momentum state and side-mode components, respectively. 
With this expansion the Hamiltonian~(\ref{eq:HBEC}) simplifies to
\begin{eqnarray}
\hat{H}_{a}^{\prime } &=&\frac{\hbar U_{0}}{2}\hat{a}^{\dagger }\hat{a}
\left( \hat{c}_{0}^{\dagger }\hat{c}_{0}+\hat{c}_{2}^{\dagger }\hat{c}
_{2}\right) +\frac{\sqrt{2}\hbar U_{0}}{4}\hat{a}^{\dagger }\hat{a}\left(
\hat{c}_{0}^{\dagger }\hat{c}_{2}+\hat{c}_{2}^{\dagger }\hat{c}_{0}\right)
\notag  \label{H side mode} \\
&+&4\hbar \omega _{r}\hat{c}_{2}^{\dagger }\hat{c}_{2},
\end{eqnarray}
where the first and second terms account for the potential energy and the
optical coupling between the zero-momentum mode and the side-mode of the
BEC respectively. The last term is the energy of the side-mode, of frequency
$4\omega _{r}=2\hbar k^{2}/ m_a$.

In the absence of particle losses we have
\begin{equation}
\hat{c}_{0}^{\dagger }\hat{c}_{0}=N-\hat{c}_{2}^{\dagger }\hat{c}_{2}\simeq
N,
\end{equation}%
where the approximate equality holds if the zero-momentum component of the
condensate is only weakly depleted by the coupling to the momentum
side-mode. Treating then that component classically via ${\hat{c}}_0$ and ${\hat{c}}_0^{\dagger }\rightarrow \sqrt{N}$, the coupling term in Eq.~({\ref{H side mode}) reduces to
\begin{equation}
\frac{\sqrt{2N}\hbar U_{0}}{4}\hat{a}^{\dagger }\hat{a}\left( \hat{c}_{2}+
\hat{c}_{2}^{\dagger }\right) \equiv \hbar \xi _{\mathrm{sm}}{\hat{a}}
^{\dagger }{\hat{a}}{\hat{Q}},
\end{equation}
where
\begin{eqnarray}
{\hat{Q}}&\equiv& \frac{1}{\sqrt{2}}({\hat{c}}+{\hat{c}}^{\dagger }),\\
\xi _{\mathrm{sm}}&\equiv& \frac{\omega _{c}}{L}\sqrt{\frac{\hbar }{m_{\mathrm{\rm sm}}4\omega _{r}}}
\end{eqnarray}
and
\begin{equation}
m_{\rm sm}=\frac{\hbar \omega _{c}^{2}}{L^{2}NU_{0}^{2}\omega _{r}}
\end{equation}
is the effective mass of the side-mode.

This indicates that the side-mode is formally analogous to a mirror whose motion is driven by radiation pressure, in full analogy to the cavity end-mirror, with position operator ${\hat Q}$ and momentum operator $\hat{P}=i(\hat{c}_{2}^{\dagger }-\hat{c}_{2})/\sqrt{2}$ satisfying the canonical commutation relation $[\hat{Q},\hat{P}]=i$. We call this effective mirror the ``sm-mirror'' in the following. (We see below that the presence of dissipation weakens somewhat the analogy between the side-mode and a moving mirror.)

A complete description of the system must include the effects of dissipation, on the optical field, the damping of the sm-mirror, and the mechanical damping of the end-mirror. These processes, which are accounted for by the Hamiltonian ${\hat H}_d$ in Eq. (\ref{H}), can be incorporated via standard quantum noise operators \cite{Tombesi2008}. They result in the coupled quantum Langevin equations
\begin{eqnarray}  \label{eq:QLE}
\frac{d\hat{a}}{dt} &=&\left( i\tilde{\Delta}+i\xi \hat{q}-i\xi _{\mathrm{sm}%
}\hat{Q}-\kappa \right) \hat{a}+\eta +\sqrt{2\kappa }\hat{a}_{\mathrm{in}},
\notag \\
\frac{d\hat{q}}{dt} &=&\omega _{m}\hat{p},  \notag \\
\frac{d\hat{p}}{dt} &=&-\omega _{m}\hat{q}+\hbar \xi \hat{a}^{\dagger }\hat{a%
}-\gamma _{m}\hat{p}+\hat{f}_{B},  \label{pp} \\
\frac{d\hat{Q}}{dt} &=&4\omega _{r}\hat{P}-\gamma _{\rm sm}\hat{Q}+\hat{f}_{1M},
\notag \\
\frac{d\hat{P}}{dt} &=&-4\omega _{r}\hat{Q}-\hbar \xi _{\mathrm{sm}}\hat{a}
^{\dagger }\hat{a}-\gamma _{\rm sm}\hat{P}+\hat{f}_{2M},  \notag
\end{eqnarray}
where $\tilde{\Delta}=\Delta _{c}-NU_{0}/2$, $\hat{a}_{\mathrm{in}}$ is the Markovian input noise of the cavity field,$\gamma _{m}$ is the mechanical energy decay rate of the end-mirror and $\hat{f}_{B}$ is the associated Brownian noise operator \cite{Pater06}. The harmonic trapping potential of the condensate, which we have ignored so far, couples the $\pm 2\hbar k$-momentum side-modes to other modes, resulting in the damping of the sm-mirror \cite{Esslinger}. It is described by the terms involving $\gamma _{\rm sm}$, with $\hat{f}_{1M}$ and $\hat{f}_{2M}$ the associated noise operators, assumed to be Markovian. We remark that unlike for physical harmonic oscillators such as the end-mirror, the damping of the sm-mirror is not simply frictional: a damping term appears now both in the Langevin equation for the momentum $\hat{P}$ and for the position $\hat{Q}$.

Both the end-mirror as well as the sm-mirror are driven by the radiation pressure of the intracavity field, which in turn depends on the positions of these oscillators. Hence the microscopic and macroscopic mechanical oscillators are coupled through the optical field, which effectively acts as a nonlinear spring connecting them.

We focus in the following on the classical dynamics of the two mirrors by treating their positions and momenta as classical variables. We also assume that the optical field decays at the fastest rate. This allows one to eliminate it adiabatically by setting its time derivative to zero, giving the steady-state intensity $|\alpha_s|^2$,
\begin{equation}
\left\vert \alpha _{s}\right\vert ^{2}=\frac{\eta ^{2}}{\kappa ^{2}+\left[
\tilde{\Delta}+\left( \frac{\xi ^{2}}{\omega _{m}}+\frac{\xi _{\mathrm{sm}%
}^{2}}{4\omega _{r}(1+\gamma _{\rm sm}^2/(4\omega_r)^2)}\right) \left\vert \alpha _{s}\right\vert ^{2}\right]
^{2}},
\end{equation}
where ${\hat a} \rightarrow \alpha$ is the classical limit.

It is known that in the absence of a condensate this system can exhibit radiation pressure optical bistability \cite{Dorsel1983}. In the present case, with effectively two moving mirrors, one finds easily that the steady-state positions of both the end-mirror, $(q_{s})$, and the sm-mirror, $(Q_{s})$ have the same bistability properties as the optical field, via the relations
\begin{eqnarray}
q_s&=&\frac{\xi |\alpha _s|^2}{\omega _m}, \nonumber \\
Q_s&=&-\frac{\xi _{\mathrm{sm}}|\alpha _s|^2}{4\omega_r(1+\gamma _{\rm sm}^2/(4\omega_r)^2)}.
\end{eqnarray}
Examples of bistability curves are shown in Fig.~\ref{bis}(a) and (b).

\begin{figure}[tbp]
\includegraphics[width=9cm]{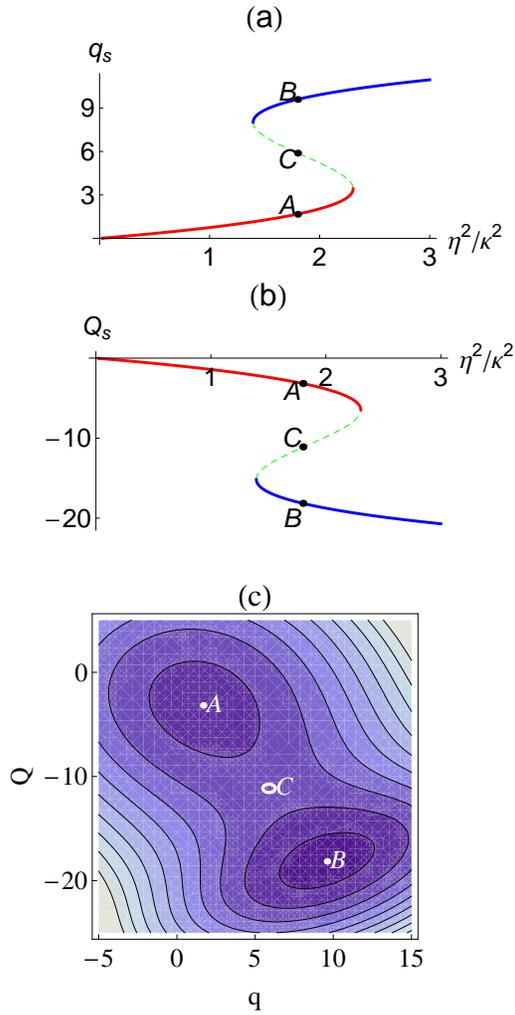}
\caption{(Color online. Bistable equilibrium position of (a) the end-mirror and (b) the sm-mirror as a function of the normalized dimensionless pump intensity $\protect\eta ^2/\protect\kappa ^{2}$. The solid lines correspond to stable solutions and the dashed lines to unstable ones. In this example the cavity length is $ L=10^{-4}\mathrm{m}$,the cavity-pump detuning $\Delta _{c}=2\protect\pi \times 15\mathrm{MHz}$, and the mass and frequency of the movable mirror are
$10^{-9}\mathrm{g}$ and $2\protect\pi \times 19\mathrm{KHz}$, respectively. The wave length of the pump field is $\lambda_p=780\mathrm{nm}$ and the decay rate of the cavity is $\kappa=2\protect\pi \times 1.3\mathrm{MHz}$, and the decay of the sm-mirror is zero. The total atom number $N=1.2\times 10^4$ and the vacuum Rabi frequency $U_0=2\protect\pi \times 3.1\mathrm{KHz}$.
(c) Two-dimensional contour plot of the potential $V\left( q,Q\right)$ at
$\protect\eta ^{2}/\protect\kappa ^{2}=1.8$, illustrating the existence of two local minima $A$ $( q=1.7,Q=-3.2) $ and $B$ $(q=9.6,Q=-18.1) $, with potential energies $E_{A}=148.8\hbar \protect\omega_m$ and $E_{B}=137.4\hbar \protect\omega_m$, respectively, and of a saddle point $C$ $(
q=5.9,Q=-11.1)$, with $E_{C}=157.6\hbar \protect\omega_m$. }
\label{bis}
\end{figure}

\section{Classical dynamics}

We assume as already indicated that the intracavity field follows adiabatically the motion of both the end-mirror and the sm-mirror, and that the retardation effects resulting from optical damping are negligible \cite{Kipp07}. Treating in addition the motion of the mirrors classically and considering time scales short enough that their mechanical damping can be ignored, their dynamics is governed by the coupled equations of motion
\begin{eqnarray}
\label{eqsofmotion}
\frac{d^{2}q}{dt^{2}} &=&-\omega _{m}^{2}q+\frac{\omega _{m}\xi \eta ^{2}}{%
\kappa ^{2}+\left( \tilde{\Delta}+\xi q-\xi _{\mathrm{sm}}Q\right) ^{2}}
\label{qq} \\
\frac{d^{2}Q}{dt^{2}} &=&-\left( 4\omega _{r}\right) ^{2}Q-\frac{4\omega
_{r}\xi _{\mathrm{sm}}\eta ^{2}}{\kappa ^{2}+\left( \tilde{\Delta}+\xi q-\xi
_{\mathrm{sm}}Q\right) ^{2}}.  \label{QQ}
\end{eqnarray}
It is easily verified that these equations can be derived from the effective classical Hamiltonian
\begin{equation}
\label{Heff}
H_{\rm eff}=T+V
\end{equation}
with
\begin{eqnarray}
T &=&\frac{\hbar \omega _{m}}{2}p^{2}+\frac{4\hbar \omega _{r}}{2}P^{2}
\notag  \label{eh} \\
V &=&\frac{\hbar \omega _{m}}{2}q^{2}+\frac{4\hbar \omega _{r}}{2}Q^{2}
\nonumber \\
&-&\frac{\hbar \eta ^{2}}{\kappa }\arctan \left[ \left( \tilde{\Delta}+\xi
q-\xi _{\mathrm{sm}}Q\right) /\kappa \right] .
\end{eqnarray}%

The Hamiltonian (\ref{Heff}) also describes the motion of a particle in the two-dimensional potential $V\left( q,Q\right) $. $V\left( q,Q\right) $ is generally a double-well potential, as illustrated in Fig.~\ref{bis}(c). The two local minima $A$ and $B$ correspond to the pairs of stable values $\{q_{A},Q_{A}\}$ and $\{q_{B},Q_{B}\}$ of Fig.~\ref{bis}(a,b), and the saddle point $C$ corresponds to the unstable values $\{q_{C},Q_{C}\}$.
The existence of such a saddle point is the source of chaotic motion in many classical models, such as e.g. the H{\' e}non-Heiles model, with the appearance of Hamiltonian chaos determined by the energy of the system \cite{Hilborn}. We can similarly expect energy-dependent transitions between regular and chaotic motion in the present case.

The dynamics of the coupled-mirrors system takes place in the four-dimensional phase space $(p,q,P,Q)$. In the absence of dissipation, energy conservation constrains the system trajectories to a three-dimensional volume of this four-dimensional space. The phase-space region occupied by the system trajectories can therefore be conveniently visualized by plotting the solutions of Eqs.~(\ref{qq}) and (\ref{QQ}) in a three-dimensional space, for instance $\left( p,q,Q\right) $. When the motion of the mirrors is regular, the trajectories are confined to closed tori in three-dimensional space that translate into regular closed orbits on Poincar{\' e} sections. When the motion becomes chaotic, these tori start to deform according to the Kolmogorov-Arnold-Moser theory \cite{Hilborn}. For chaotic motion the
trajectories have no special geometric shape, and the Poincar{\' e} sections turn into a ``chaotic sea'' without any distinctive structure.

\begin{figure}[tbp]
\includegraphics[width=8cm]{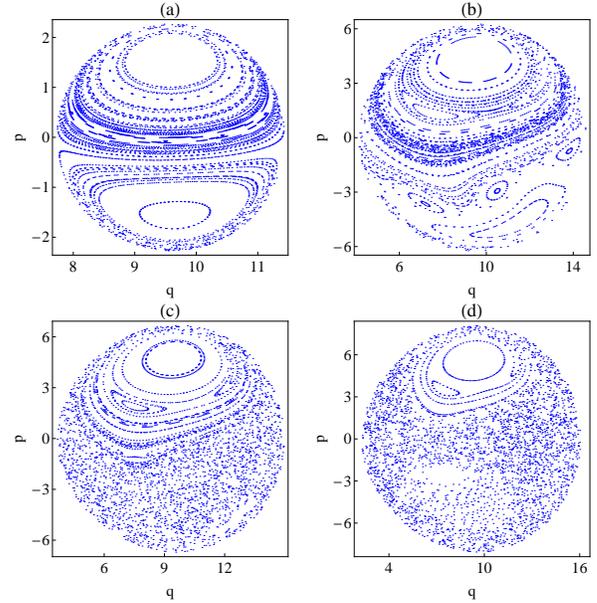}
\caption{(Color online). Poincar{\' e} sections $(q,p)$ of the coupled-mirrors system for (a) an energy $E=140\hbar \omega_m$ close to the local minimum energy $E_B$, (b) an energy $E=157\hbar \omega_m$ near but below the saddle-point energy $E_C$, (c) $E=160\hbar \omega_m$ above $E_C$, and (d) $E=170\hbar \omega_m$ significantly above $E_C$. Same parameters as in Fig.~2.}
\label{poin}
\end{figure}

In the following we concentrate for concreteness on Poincar{\' e} sections in the $(p,q)$ plane for $Q=Q_{B}$ and $P>0$. When solving Eqs.~(\ref{qq}) and ~(\ref{QQ}) numerically for random initial conditions corresponding to a total energy slightly above $E_{B}$ but much lower than the saddle-point energy $%
E_{C}$, we observe that the Poincar{\' e} section is characterized by a series of regular elliptic orbits, see Fig.~\ref{poin}(a). These orbits can be divided into two families that surround two points ($p>0$ and $p<0$) on the line $q=q_{B}$. The oscillations of the end-mirror and the sm-mirror are in
phase for $p>0$ and out of phase for $p<0$.  The observations imply that at low energies the motion of the coupled mirrors is quasi-periodic.  We note that similar Poincar{\' e} sections have previously been observed in the case of integrable two-dimensional potentials with anharmonicities up to terms of fourth order \cite{Cleary90}, as well as for a double pendulum at low energy \cite{Korsch08}. In the effective particle analogy, the particle is trapped in potential well $B$.

The situation starts to change when the total energy of the system is close to $E_{c}$. We then find that some of the out-of-phase elliptic orbits break into a series of elliptical islands, see Fig.~\ref{poin}(b). This is an indication that the closed torii start to be destroyed and the motion is tending to chaos. When the energy of the system is slightly larger than the saddle-point energy, as shown in Fig.~\ref{poin}(c), all out-of-phase elliptic orbits dissolve into a chaotic sea, although some of the in-phase elliptic orbits still survive. Finally, when the energy is increased still further, those surviving orbits also degenerate into islands or dissolve in the chaotic sea, see Fig.~\ref{poin}(d).

\begin{figure}[tbp]
\includegraphics[width=9cm]{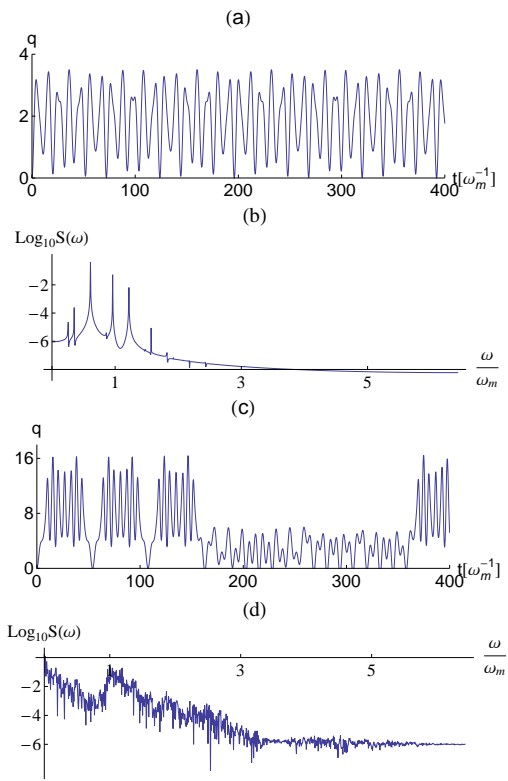}
\caption{(a) $q$ as a function of time for the initial condition $
p(0)=q(0)=P(0)=Q(0)=0$ and the input intensity $\eta ^{2}/
\kappa ^{2}=1.8$. (b) The power spectrum of $q(t)$, illustrating the quasi-periodic motion of the cavity end-mirror. The chaotic motion of the
end-mirror is displayed by (c) and (d) which are for the same parameters as (a) and (b), except that $\eta ^{2}/\kappa ^{2}=2$. All other parameters as in Fig.~\ref{bis}.}
\label{spec}
\end{figure}

Figure~\ref{spec} shows the motion and the power spectrum of the end-mirror in both the quasi-periodic and the chaotic regimes. The series of Lorentzian-shaped peaks in the power spectrum of Fig.~\ref{spec}(b) is a clear signature of quasi-periodic motion, and should be contrasted to the broad, structureless spectrum of Fig.~\ref{spec}(d), characteristic of the chaotic motion of $q(t)$ shown in Fig.~\ref{spec}(c). (The oscillations of $Q(t)$, not shown in Fig.~\ref{spec}, are of course likewise chaotic.).
In practice the chaotic dynamics of the system can be observed by detecting
the light transmitted or reflected by the cavity. This light carries
experimental signatures both of the end-mirror motion \cite{Carmon05,Carmon07} and of the sm-mirror dynamics \cite{Esslinger}.

In currently experimentally realizable systems the mechanical decay of the end-mirror is typically of the order of perhaps $\gamma _{m}\sim 1 \mathrm{Hz}$. A soft harmonic trap for the condensate, with trapping frequency $2\pi \times 1\mathrm{Hz}$ gives likewise a sm-mirror damping rate of the order of $\gamma _{sm}\sim 1 \mathrm{Hz}$. In addition, one needs to account for the condensate lifetime, limited primarily by three-body collisions. For $^{87}{\mathrm{Rb}}$ atoms the three-body loss rate constant has been
measured to be $K_{3}=5.8\times 10^{-30}\mathrm{cm}^{\mathrm{6}}\mathrm{Hz}$
\cite{Burt97}. For an atomic density of $n\sim 10^{13}
\mathrm{cm}^{\mathrm{-3}}$ and $N \sim 1.2 \times 10^{4}$, the three-body loss rate is therefore $K_{3}n^{2}N\sim 1 \mathrm{Hz}$. Hence the motion of the system is justifiably described in the framework of Hamiltonian chaos for time scales shorter than a second.

For larger decay rates (or longer times), there is in principle the possibility that the system undergoes dissipative chaos. To determine whether this is the case we have carried out a linear stability analysis of Eqs.~(\ref{eqsofmotion}), properly generalized to include the effects of mirror damping. We have found numerically that the system is always stable in the bad cavity limit considered here. The situation is different in the case of a good (high finesse) cavity, where the intracavity field can no longer be adiabatically eliminated. In this case, the system can enter a regime of dissipative chaos, consistent with previous studies of chaos in optomechanical systems with a single moving mirror \cite{Carmon05,Carmon07}. Figure~\ref{nochaos} shows an example of mirror dynamics in the presence of larger mirror damping. It illustrates the disappearance of the chaotic motion and the approach to a stable steady state resulting from the damping guiding the system toward a a local energy minimum.

\begin{figure}[tbp]
\includegraphics[width=7cm]{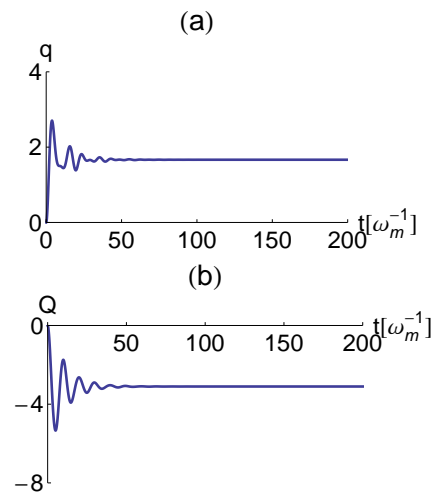}
\caption{ The same parameters as those in Fig.\protect\ref{spec}, but with
decay terms, $\protect\gamma_m=\protect\gamma_c=0.1 \protect\omega_m$ (a)
The position of the mirror as a function of time. (b) The virtual positon of
the sm-mirror as a function of time. }
\label{nochaos}
\end{figure}

\section{Conclusion and outlook}

In conclusion, we have described the dynamics of a hybrid optomechanical
system where a BEC is strongly coupled to the intracavity optical field of a Fabry-P{\' e}rot cavity with a movable end-mirror. The photon recoil associated with the virtual absorption and emission of light by the condensate results in the formation of an effective side-mode mirror that is driven just like the end-mirror by radiation pressure. In addition the light field couples nonlinearly this effective mirror and the end-mirror. We have shown that in the classical limit this optomechanical system displays radiation pressure induced bistability. In the bad cavity limit, and for times short enough that mirror damping can be safely ignored, the coupled dynamics of the mirrors can be described by an effective double-well potential. For appropriate initial conditions the system is then characterized by the onset of Hamiltonian chaos.
  
Future work will extend this discussion to the good cavity limit, where the onset of dissipative chaos can also be expected. We will also consider a full quantum mechanical description of the mirrors motion, and study the optically induced quantum correlations and entanglement between the sm-mirror and the end-mirror. Additional goals include the quantum control of the mirror motion by the atoms, and conversely of the atoms by the optomechanical system.

\acknowledgements
We thank S. Singh and D. Goldbaum for useful discussions. This work is supported by the U.S. Office of Naval Research, the U.S.
National Science Foundation, and the U.S. Army Research Office.


\begin{thebibliography}{99}
\bibitem{note} This Hamiltonian chaos should of course not be confused with dissipative chaos previously discussed in optomechanical setups \cite{Carmon05,Carmon07}.
\end{thebibliography}
\end{document}